\documentclass{PoS}
\usepackage{hepunits}
\newcommand{\pb}{\textrm{~pb}}
\usepackage{pstricks}
\usepackage{epsf}

\title{Timelike Compton Scattering from JLAB to RHIC and LHC energies}
\ShortTitle{Timelike Compton Scattering from JLAB to RHIC and LHC energies}

\author{
B.~Pire$^a$, L.~Szymanowski$^b$ and \speaker{J.~Wagner$^b$} \\
\llap{$^a$}CPHT, {\'E}cole Polytechnique, CNRS, 91128 Palaiseau, France\\
\llap{$^b$}National Center for Nuclear Research, Warsaw\\
E-mail: 
\email{Bernard.Pire@cpht.polytechnique.fr},
\email{lech.szymanowski@ncbj.gov.pl},
\email{jakub.wagner@ncbj.gov.pl}
}

\abstract{Timelike Compton scattering (TCS) i.e. the exclusive photoproduction of a lepton pair with large invariant mass nicely complements the already successful experimental study of deeply virtual Compton scattering (DVCS). The same Generalized Parton Distributions enter both amplitudes, which offer a promissing way to access the quark and gluon nucleon structure.  We review recent progress in this domain, emphasizing the fact that analyticity and factorization properties dictate the relation of the NLO corrections to TCS to those of DVCS.  We also stress that data on TCS at high energy should be available soon thanks to the proposed experimental program at JLab at 12 GeV, and that, before the future high energy electron ion colliders become reality, the study of ultraperipheral collisions at the RHIC and LHC may open a window on quark and gluon GPDs at very small skewness.
.}

\FullConference{Sixth International Conference on Quarks and Nuclear Physics\\
        April 16-20, 2012\\
        Ecole Polytechnique, Palaiseau, France}

\begin{document}

\section{Intoduction}
Almost two decades after its first stages~\cite{Muller:1994fv}, the study of deeply virtual Compton scattering (DVCS),
 i.e., $\gamma^* p \to \gamma p$, and more generally of hard exclusive reactions in a generalized Bjorken regime, has now entered a phase where many theoretical and experimental progresses can merge to enable a sensible extraction of generalized parton
distributions (GPDs). 
Indeed, the measurement of GPDs should contribute in a decisive way to
our understanding of how quarks and gluons build hadrons~\cite{gpdrev}. In particular the transverse
location of quarks and gluons become experimentally measurable via the transverse momentum dependence of the GPDs \cite{Burk}.

Timelike Compton scattering (TCS) \cite{TCS}  $$\gamma(q) N(p) \to \gamma^*(q') N(p') \to l^-(k) l^+(k') N(p')$$
   at small $t = (p'-p)^2$ and large \emph{timelike} virtuality $(k+k')^2=q'^2 = Q^2$ of the final state
 dilepton, shares many features with its ``inverse'' process, DVCS. The Bjorken variable in the TCS case is $\tau = Q^2/s $
 with $s=(p+q)^2$. One also defines $\Delta = p' -p$  ($t= \Delta^2$) and the scaling variable $xi$ and skewness $eta$:
$\xi  = - \frac{(q+q')^2}{2(p+p')\cdot (q+q')} \,\approx\,
           \frac{ - Q^2}{2s  - Q^2} $, 
$\eta = - \frac{(q-q')\cdot (q+q')}{(p+p')\cdot (q+q')} \,\approx\,
           \frac{ Q^2}{2s  - Q^2}$
where the approximations hold in  the extended Bjorken regime, where masses and $-t$ are small with respect of $Q^2$ ($s$ is always larger than $Q^2$ ). 
\begin{figure}[htb]
  \centering
  \includegraphics[width=0.25\textwidth]{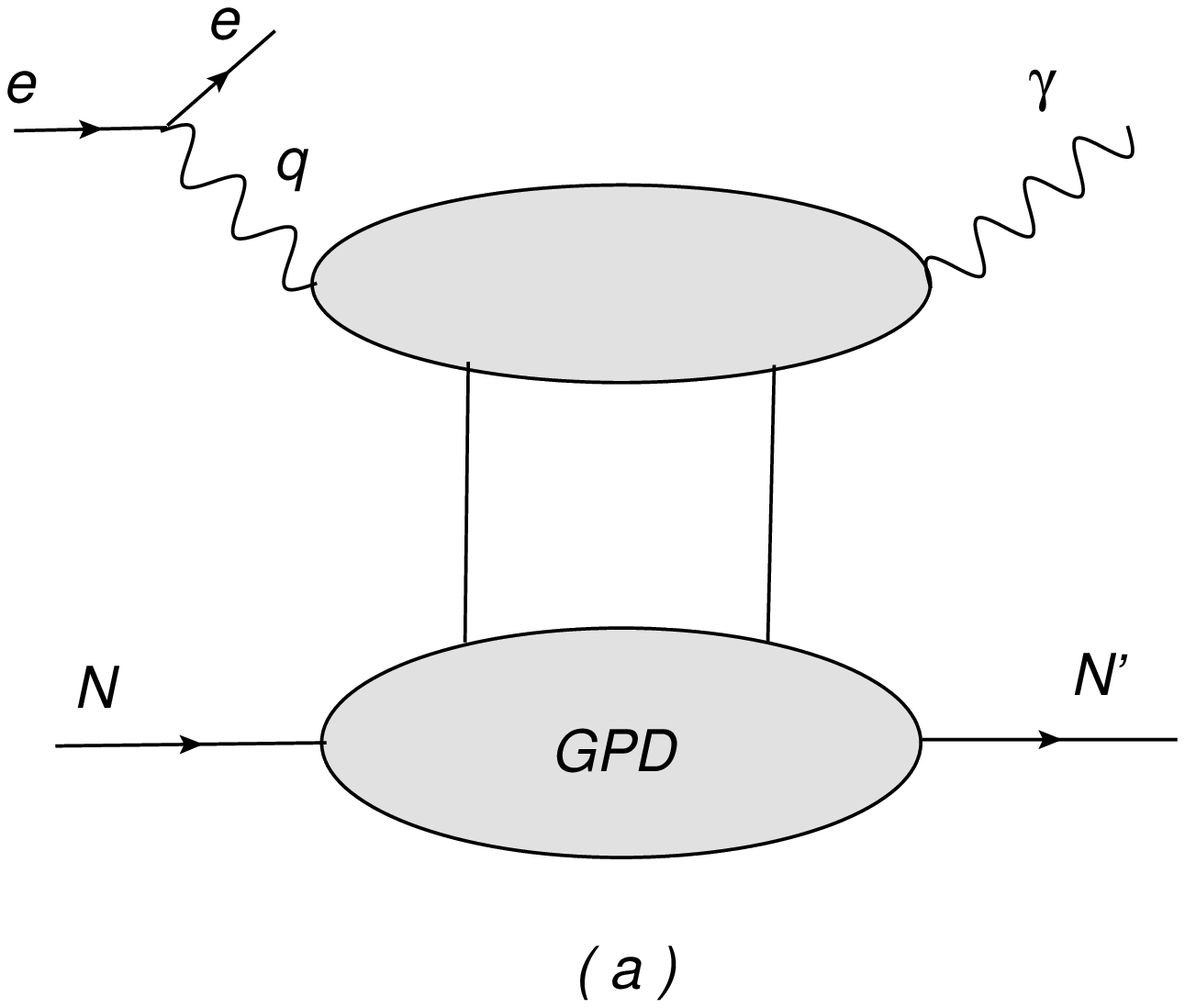}
  \includegraphics[width=0.25\textwidth]{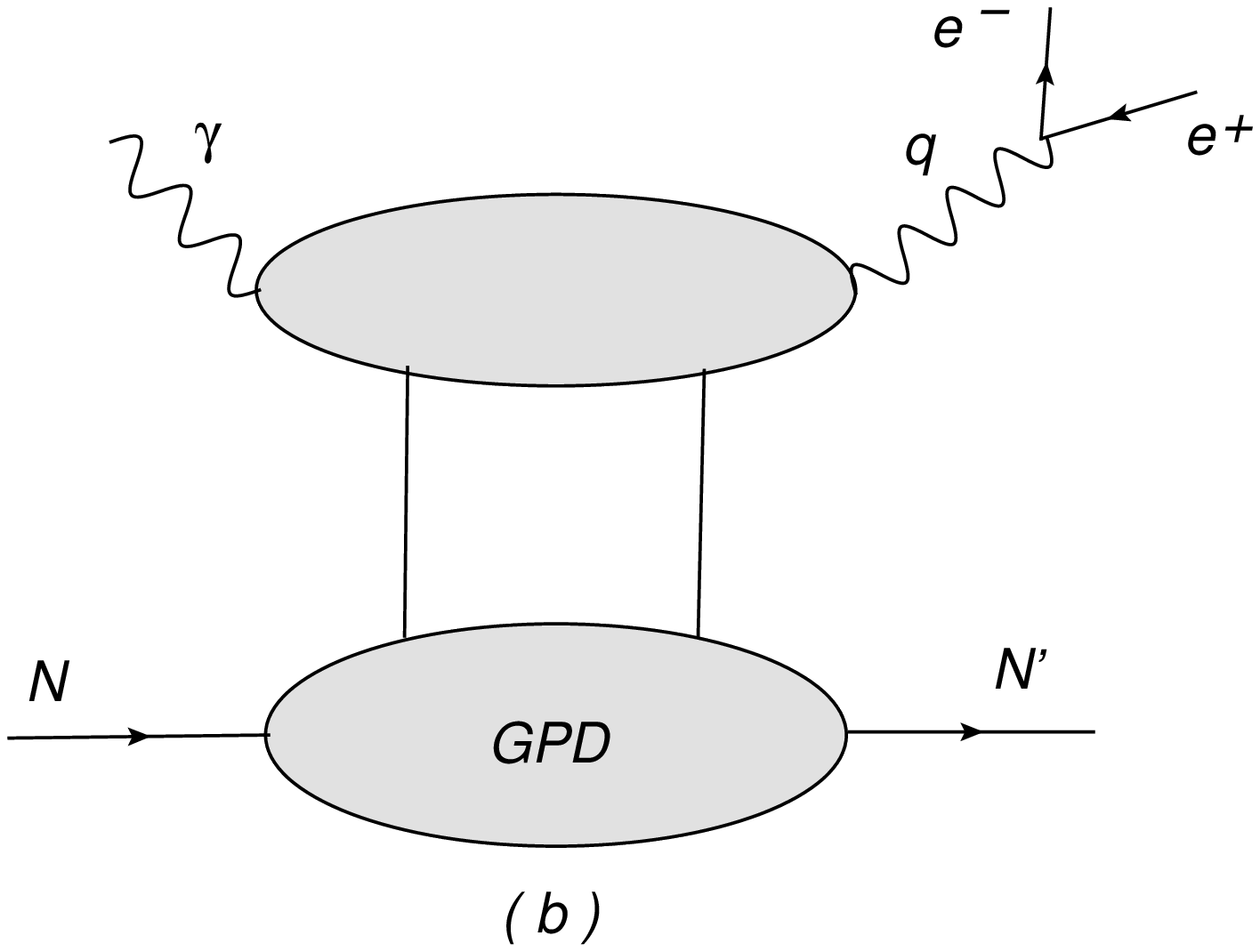}
  \caption{(a) Deeply Virtual Compton Scattering and (b) Timelike Compton Scattering}
  \label{Fig:DVCSTSC}
\end{figure}
\section{Basic properties and first experimental results}
In the region where the final photon virtuality is large, the Compton amplitude is given by the convolution of hard scattering coefficients, calculable in perturbation theory, and generalized parton distributions, which describe the nonperturbative physics of the process. The physical process where to observe TCS, is photoproduction of a heavy lepton pair, $$\gamma N \to \mu^+\!\mu^-\, N ~~~~~or~~~~~ \gamma N \to e^+\!e^-\, N\;.$$
A QED process, the Bethe-Heitler (BH)
mechanism  $\gamma(q)  \gamma^*(-\Delta)  \to l^-(k) l^+(k') $ contributes at the amplitude level. 
This latter process has a very peculiar angular dependence and overdominates the TCS process if
one blindly integrates over the final phase space. One may however choose kinematics where 
the amplitudes of the two processes are of the same order of magnitude, and  use specific observables sensitive to the interference of the two amplitudes. Since the amplitudes for the Compton and Bethe-Heitler
processes transform with opposite signs under reversal of the lepton
charge,  it is possible to project out the interference term through a clever use of the angular distribution of the lepton pair \cite{TCS}.

 First attempts to measure TCS, and to confront the theoretical predictions with data were already performed at JLab at 6 \GeV \cite{Rafayel}, and may serve as a feasibility test for a proposed experiment with higher energy after upgrade to 12 \GeV.

\section{TCS at next to leading order}
\label{sec:NLO}
After proper renormalization, the Compton scattering amplitude reads in its factorized form:
\begin{eqnarray}
\mathcal{A}^{\mu\nu} &=& -g_T^{\mu\nu}\int_{-1}^1 dx 
\left[
\sum_q^{n_F} T^q(x) F^q(x)+T^g(x) F^g(x)
\right] \nonumber \\
&+& i\epsilon_T ^{\mu\nu}\int_{-1}^1 dx 
\left[
\sum_q^{n_F} \tilde{T}^q(x) \tilde{F}^q(x)+\tilde{T}^g(x) \tilde{F}^g(x)
\right] \,,
\label{eq:factorizedamplitude}
\end{eqnarray}
where renormalized coefficient functions for the vector case are given by:
\begin{eqnarray}
T^q(x)&=& \left[ C_{0}^q(x) +C_1^q(x) +\ln\left(\frac{Q^2}{\mu^2_F}\right) \cdot C_{coll}^q(x)\right] - ( x \to -x )  \,,\nonumber\\
T^g(x) &=& \left[ C_1^g(x) +\ln\left(\frac{Q^2}{\mu^2_F}\right) \cdot C_{coll}^g(x)\right] +( x \to -x )
\,.
\label{eq:ceofficients}
\end{eqnarray} 
and similarily (but with different symmetry in $x$) for the axial quantities $\tilde{T}^q, \tilde{T}^g$. Results for TCS \cite{Pire:2011st} are connected to the well-known DVCS results \cite{BelMueNieSch00}, through a simple relation \cite{MPSW}: 
\begin{eqnarray}
^{TCS}T(x) = \pm \left(^{DVCS}T(x) +  i \pi C_{coll}(x)\right)^* \,,
\label{eq:TCSvsDVCS}
\end{eqnarray}
where +(-) sign corresponds to vector (axial) case. 
\begin{figure}[htb]
  \centering
  \includegraphics[width=0.35\textwidth]{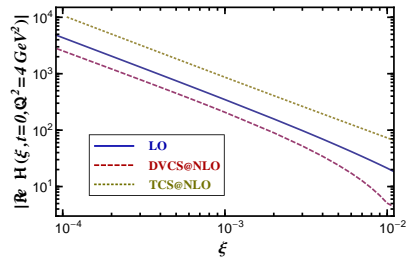}
  \caption{The real part of CFF $\mathcal{H}$ vs.~$\xi$ with $\mu^2=Q^2= 4 \textrm{~GeV}^2$  and $t=0$ at LO (solid) and NLO  for DVCS (dashed). For  TCS at NLO  its negative value is shown as dotted curve.  }
  \label{fig:NLO}
\end{figure}
\begin{figure}[htb]
  \centering
 \includegraphics[width=0.30\textwidth]{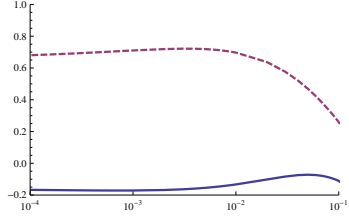}
 \caption{ The ratio of the real to the imaginary part of CFF ${\mathcal{H}}$ in TCS at LO (solid) and NLO (dashed) as a function of $\xi$ with $\mu^2=Q^2= 4 \textrm{~GeV}^2$  and $t=0$	.}
  \label{fig:NLO2}
\end{figure}
Analyticity of the factorized amplitude is the basic property that allows us to derive this new relation. Analyticity, which is a consequence of causality in relativistic field theory, and factorization of short distance vs long distance properties, are common tools in many fields of theoretical physics. Our instance is to our knowledge the first case where they are put together to obtain useful relations between observables.

It is convenient to express the amplitude in terms of Compton From Factors (CFFs) defined as a convolution of coefficient functions with GPD's. For example, in unpolarized case the amplitude is given by:
\begin{equation}
\mathcal{A}^{\mu\nu} = - e^2 \frac{1}{(p+p')^+}\, \bar{u}(p^{\prime}) 
\left[\,
   g_T^{\mu\nu} \, \Big(
      {\mathcal{H}} \, \gamma^+ +
      {\mathcal{E}} \, \frac{i \sigma^{+\rho}\Delta_{\rho}}{2 M}
   \Big)
\,\right] u(p) \,,
\label{eq:amplCFF}
\end{equation}
and Compton form factors ${\mathcal{H}}$ and ${\mathcal{E}}$:
\begin{eqnarray}
\mathcal{H}(\xi,\eta,t) &=& - \int_{-1}^1 dx \,
\left(\sum_q T^q(x,\xi,\eta)H^q(x,\eta,t)
 + T^g(x,\xi,\eta)H^g(x,\eta,t)\right) \nonumber \\
\mathcal{E}(\xi,\eta,t) &=& - \int_{-1}^1 dx \,
\left(\sum_q T^q(x,\xi,\eta)E^q(x,\eta,t) 
+T^g(x,\xi,\eta)E^g(x,\eta,t)\right)
\label{eq:CFF}
\end{eqnarray}

The NLO relation (\ref{eq:TCSvsDVCS}) tells us, that if scaling violations are small, the Compton From Factors  and their timelike verison (TFFs) can be obtained from each other by complex conjugations. Moreover, GPD model studies indicate that in the valence region, i.e., for $\xi \sim 0.2$,   CFFs  might only evolve mildly. This rather generic statement, which will be quantified by model studies \cite{Moutarde}, might be  tested in future (after 12GeV upgrade) Jefferson Lab experiments. On the other hand we expect huge  NLO corrections
to $\Re{\rm e}{^{TCS}{\cal H}} \stackrel{\rm LO}{=} \Re{\rm e}{^{DVCS}{\cal H}}$, induced by $\Im{\rm m}{\cal H}$. DVCS and TCS have rather similar effects on the imaginary parts, diminishing its absolute value. The situation is very different for the real part where we observe huge differences between NLO DVCS and NLO TCS corrections.
Utilizing Goloskokov-Kroll model
for $H$ GPDs \cite{Goloskokov:2006hr}, we illustrate this effect
in Fig.~\ref{fig:NLO} for $10^{-4}\le  \xi \le 10^{-2}$, accessible in a suggested Electron-Ion-Collider \cite{Boer:2011fh,LHeC}, and $t=0$.  We plot $\Re{\rm e}{\cal H}$ vs.~$\xi$, for LO DVCS or TCS (solid), NLO DVCS (dashed) and NLO TCS (dotted) at
the input scale $\mu^2={\cal Q}^2 = 4 \textrm{~GeV}^2$. In the case of NLO TCS $-\Re{\rm e}{^T{\cal H}}$ is shown, since even the sign changes. We read off that the NLO correction to $\Re{\rm e}{^T{\cal H}}$ is of the order of $-400\%$ and so
the real part in TCS becomes of similar importance as the imaginary part. This fact is also illustrated by Fig.~\ref{fig:NLO2}, where we show the ratio of the real to the imaginary part of CFF ${\mathcal{H}}$ in TCS at LO and NLO as a function of $\xi$ in the same model and values of $\mu^2$ and $Q^2$.

This NLO prediction is testable via a
lepton-pair angle asymmetry, governed by  $\Re{\rm e}{^T{\cal H}}$ \cite{TCS}. 
\section{Ultraperipheral collisions}
\begin{figure}
\begin{center}
\epsfxsize=0.39\textwidth
\epsffile{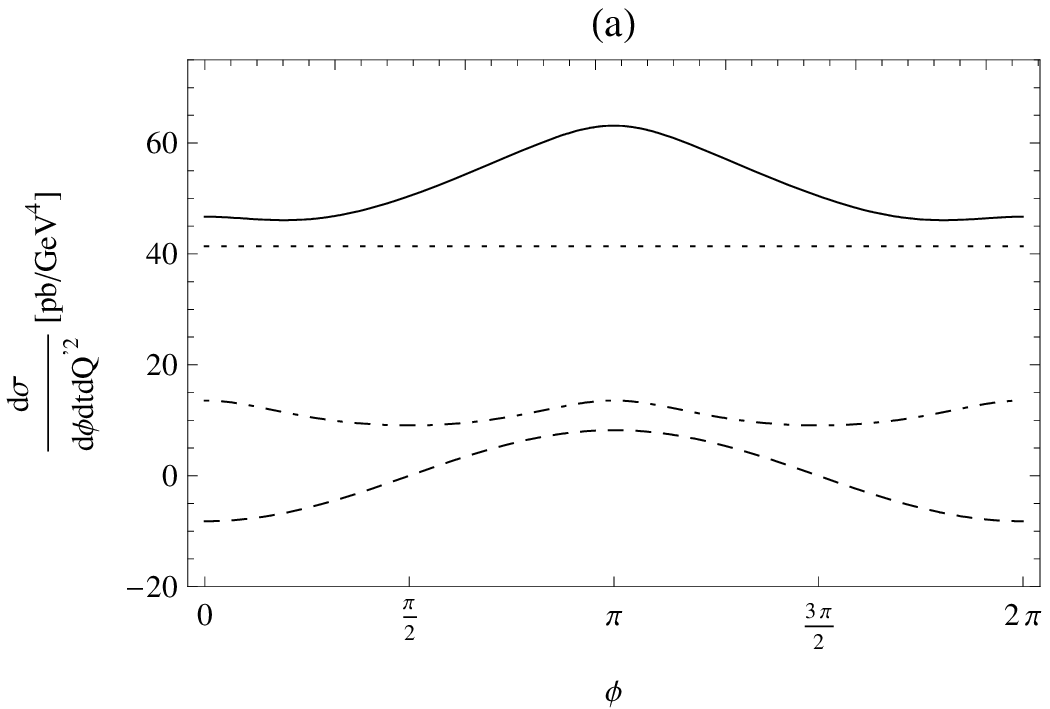}
\hspace{0.05\textwidth}
\epsfxsize=0.39\textwidth
\epsffile{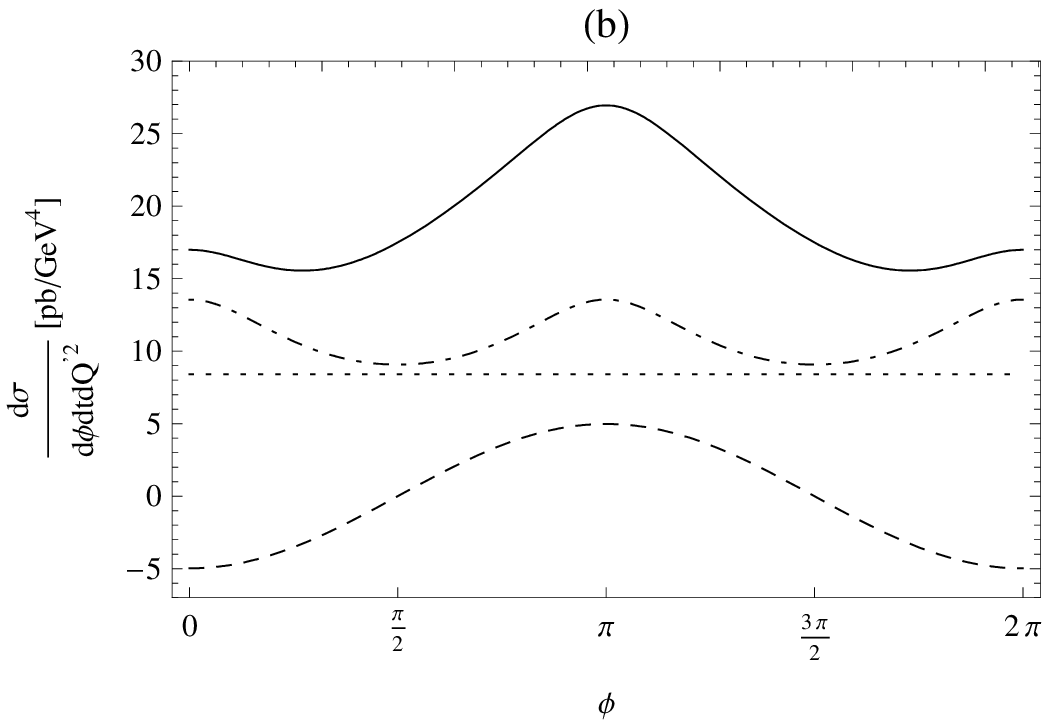}
\caption{
The differential cross sections (solid lines) for $t =-0.2 \GeV^2$, ${Q'}^2 =5 \GeV^2$ and integrated 
over $\theta = [\pi/4,3\pi/4]$, as a function of $\varphi$, for $s=10^7 \GeV^2$ (a), 
$s=10^5 \GeV^2$(b) with $\mu_F^2 = 5 \GeV^2$. We also display  the
Compton (dotted), Bethe-Heitler (dash-dotted) and Interference (dashed) contributions. 
}
\label{Interf}
\end{center}
\end{figure}

In Fig. \ref{Interf} we show the interference contribution to the cross section in comparison to the Bethe Heitler and Compton processes, for various values of photon proton  energy 
squared $s = 10^7 \GeV^2$ and $s=10^5 \GeV^2$. We observe that for 
larger energies the Compton process dominates, whereas for $s=10^5 \GeV^2$ all contributions are comparable. 

As described in \cite{BeKlNy} the cross section for photoproduction in hadron collisions is given by:
\begin{equation}
\sigma_{pp}= 2 \int \frac{dn(k)}{dk} \sigma_{\gamma p}(k)dk \,,
\end{equation}
where $\sigma_{\gamma p} (k)$ is the cross section for the 
$\gamma \,p \to p\, l^+ l^-$ process and $k$ is the photon energy. 
$\frac{dn(k)}{dk}$ is an equivalent photon flux (the number of photons with energy $k$).
In Ref. \cite{PSW1} we analyzed the possibility to measure TCS at the LHC. The pure Bethe - Heitler contribution to $\sigma_{p p}$, integrated over  $\theta = [\pi/4,3\pi/4]$, $\phi = [0,2\pi]$, $t =[-0.05 \GeV^2,-0.25 \GeV^2]$, ${Q'}^2 =[4.5 \GeV^2,5.5 \GeV^2]$, and photon energies $k =[20,900]\GeV $  gives $\sigma_{pp}^{BH} = 2.9 \pb $. The Compton contribution (calculated with NLO GRVGJR2008 PDFs, and $\mu_F^2 = 5 \GeV^2$) gives $\sigma_{pp}^{TCS} = 1.9 \pb$.

We have choosen the range of photon energies in accordance with expected capabilities to tag photon energies
at the LHC. This amounts to a large rate of order of $10^5$ events/year at the LHC with its nominal 
luminosity ($10^{34}\,$cm$^{-2}$s$^{-1}$). 
\begin{figure}[htb]
  \centering
  \includegraphics[width=0.35\textwidth]{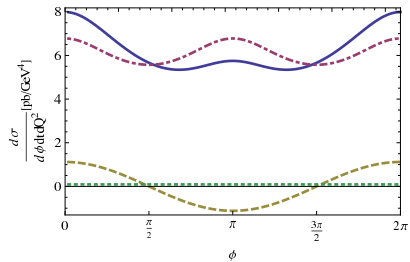}
  \caption{Total (solid), TCS (dotted), BH (dash-dotted) and intereference (dashed) differential cross section for ultraperipheral collisions at RHIC.}
  \label{Fig:RHIC}
\end{figure}
Figure \ref{Fig:RHIC} shows predictions obtained for ultraperipheral collisions at RHIC, using KG model for $t=-0.1 \GeV^2$ and $\sqrt{s_{pp}}=500 \GeV^2$. Only BH contribution gives $10^3$ events for $10^7 s$.

\section{Conclusions}
In conclusion, we advocated that timelike Compton scattering is a reaction with many opportunities, both at current and future lepton  facilities and in  hadron colliders thanks to the ultraperipheral reactions where hadron beams give birth to intense photon beams. The perturbative analysis of the coefficient functions for both DVCS and TCS is becoming more and more under control, and resummation strategies \cite{Altinoluk:2012fb} are now undertaken. A trustful extraction of generalized parton distributions from present and future data will benefit from these progresses (see i.e. \cite{Sabatie,Moutarde:2012ht}).

\section*{Acknowledgements}
This work is partly supported by the Polish Grant NCN No DEC-2011/01/D/ST2/02069 and the Joint Research Activity "Study of Strongly Interacting Matter" (acronym HadronPhysics3, Grant Agreement n.283286) under the
Seventh Framework Programme of the European Community.

\end{document}